\documentclass[prl,twocolumn,showpacs,preprintnumbers,amsmath,amssymb,superscriptaddress]{revtex4}

\usepackage{graphicx}
\usepackage{dcolumn}
\usepackage{bm}

\def \ed {\end{document}}
\def\Fbox#1{\vskip1ex\hbox to 8.5cm{\hfil\fboxsep0.3cm\fbox{%
  \parbox{8.0cm}{#1}}\hfil}\vskip1ex\noindent}  

\def\be{\begin{equation}}\def\ee{\end{equation}}
\def\bea{\begin{eqnarray}}\def\eea{\end{eqnarray}}
\def\bse{\begin{subequations}}\def\ese{\end{subequations}}
\newcommand{\BE}[1]{\begin{equation}\label{#1}}
\newcommand{\BEA}[1]{\begin{eqnarray}\label{#1}}
\newcommand{\BSE}[1]{\begin{subequations}\label{#1}}

\let \= \equiv \let\*\cdot \let\~\widetilde \let\^\widehat \let\-\overline

\def\<{\left\langle}    \def\>{\right\rangle}
\def\({\left(}          \def\){\right)}
 \def \[ {\left [} \def \] {\right ]}

\makeatletter
\AtBeginDocument{\@ifpackageloaded{natbib}{\ifNAT@numbers\if@filesw\immediate\write\@auxout{\string\global\string\NAT@numberstrue}\fi\fi}{}}
\makeatother

\begin{document}

\preprint{APS/123-QED}

\title{Bogoliubov-wave turbulence in Bose-Einstein condensates} 

\author{Kazuya Fujimoto}
\affiliation{Department of Physics, Osaka City University, Sumiyoshi-ku, Osaka 558-8585, Japan}

\author{Makoto Tsubota}
\affiliation{Department of Physics, Osaka City University, Sumiyoshi-ku, Osaka 558-8585, Japan}
\affiliation{The OCU Advanced Research Institute for Natural Science and Technology (OCARINA), Osaka City University, Sumiyoshi-ku, Osaka 558-8585, Japan}

\date{\today}

\begin{abstract}
We theoretically and numerically study Bogoliubov-wave turbulence in three-dimensional atomic Bose-Einstein condensates with the Gross-Pitaevskii equation, investigating 
three spectra for the macroscopic wave function, the density distribution, and the Bogoliubov-wave distribution.
In this turbulence, Bogoliubov waves play an important role in the behavior of these spectra, so that we call it  Bogoliubov-wave turbulence. 
In a previous study [D. Proment \textit{et al.}, Phys. Rev. A \textbf{80}, 051603(R) (2009)], a $-3/2$ power law in the spectrum for the macroscopic wave function was suggested by using  weak wave turbulence theory, but we find that another $-7/2$ power law appears in both theoretical and numerical calculations. Furthermore, we focus on the spectrum for the density distribution, which can be observed in experiments, discussing the possibility of  experimental observation. Through these analytical and numerical calculations, we also demonstrate that the previously neglected condensate dynamics induced by the Bogoliubov waves is  remarkably important. 
\end{abstract}


\maketitle

\section{Introduction}
In nature, complex and chaotic flow universally appears in various systems from small to large scale, for instance,   in 
watercourses, rivers, oceans, and the atmosphere. 
Such a flow,  called turbulence, has attracted considerable attention in various fields
because understanding it leads to significant development of fundamental and applied sciences such as mathematics, physics, and technology.
Though turbulence has been studied for a long time, however, it is considered to be one of the unresolved problems
and becomes an important issue in modern physics. 
 
Originally, turbulence has been studied in classical fluids, where it is important to investigate statistical quantities such as the probability density function of the velocity field  \cite{Davidson,Frisch}. 
One of the most significant quantities is the kinetic energy spectrum corresponding to a two-point correlation function 
of the velocity field, which is known to exhibit the famous Kolmogorov $-5/3$ power law. 
This Kolmogorov law appears as a result of the constant-flux energy cascade in  wave number space, 
and the observation of this power law is very important in the study of turbulence. 

In quantum fluids, turbulence also appears and has been investigated for many years \cite{Hal,Vinen,Skrbek12,White13,Barenghi14}.
This turbulence is called quantum turbulence (QT), in contrast to classical turbulence (CT), and was originally developed in the field of superfluid helium. 
QT offers rich phenomena such as thermal counterflow \cite{Vinen57,Tough}, Richardson cascade \cite{Mauer,Nore,KT05}, Kelvin wave cascade \cite{Kozik04,Lvov10}, the bottleneck effect \cite{Lvov07}, velocity statistics \cite{Paoletti08J}, and the reconnection of quantized vortices \cite{Paoletti08}, leading to a deep understanding of turbulence, and it has been actively studied until today. 

Recently, atomic Bose-Einstein condensates (BECs) have been providing a novel stage for turbulence studies in quantum fluids 
\cite{Hal, White13, Barenghi14}.
This system has three advantages for investigating turbulence in quantum fluids. The first advantage is the experimental visualization of
the  density and  vortex distribution \cite{Neely10}, which gives us a large amount of information about turbulence. Such an observation is difficult in superfluid helium, though the vortices have been visualized  recently by use of tracer particles \cite{Paoletti08J,Paoletti08}. 
The second advantage is the high controllability of the system by optical techniques \cite{PS}, which enable us to change the shape of the quantum fluid and the interaction between particles. This can access new problems: the geometrical dependence of turbulence such as two-dimensional QT. This kind of study is being actively investigated. 
The third advantage is the realization of multicomponent BECs such as binary \cite{KTU} and spinor BECs \cite{KU,stamper}. 
The hydrodynamics in this kind of quantum fluid will give us novel physics, and investigations  in this area have recently begun \cite{Gautam10,Sasaki09,Takeuchi10,Karl13,Koba14,Vill14,FT14}. 
Thus, turbulence in the atomic BECs can address exotic phenomena not found in superfluid helium. 

At present, there is some research on  turbulence in atomic BECs, and 
most  theoretical and numerical studies consider turbulence with many quantized vortices. 
In three- and two-dimensional systems, when many quantized vortices are nucleated, the Kolmogorov $-5/3$ power law 
is numerically confirmed in incompressible kinetic energy \cite{Parker,KT07,Gou09, Bradley12,Reeves13}. 
Recently,  two-dimensional QT has been actively investigated, with discussion of features characteristic of  two-dimensional systems such as inverse energy cascades \cite{Gou09,Bradley12,Reeves13}, vortex clustering \cite{White12,Reeves13,Simula14}, negative temperature \cite{Simula14}, and decay of vortex number \cite{Stagg15}.

In experiments, some groups have succeeded in generating turbulence in  atomic BECs \cite{Henn09,Reeves12,Neely13,Kwon14}, where 
it has been possible to obtain  turbulence with many quantized vortices. 
All these experimental studies focus on quantized vortices, investigating the anomalous expansion \cite{Henn09}, vortex dynamics \cite{Reeves12}, annihilation of vortices \cite{Kwon14}, and so on.
However, the kinetic-energy spectrum corresponding to a two-point correlation function of the velocity field has not yet been observed, and we 
cannot confirm the Kolmogorov $-5/3$ power law. This means that Kolmogorov turbulence has not been observed. 

Let us elucidate the meaning of Kolmogorov turbulence. In the usual hydrodynamic turbulence (HT), the vortices give important structures, whose 
dynamics generates the $-5/3$ power law in the kinetic-energy spectrum due to a constant-flux cascade for the kinetic energy. 
However, in other systems, various power laws appear as a result of the constant-flux cascade for some quantity such as the wave 
energy, wave action, and so on. 
In these cases, the common physics for the appearance of the power law is the constant-flux cascade. 
In this paper, we call such turbulence with the constant-flux cascade Kolmogorov turbulence. 
In the turbulence study, it would be very important to observe Kolmogorov turbulence; however, observing the superfluid velocity directly in turbulence of atomic BECs is difficult, so we cannot confirm whether Kolmogorov turbulence appears  as of now.

To shed light on the above problem, we focus on weak wave turbulence (WT) with a strong condensate in atomic BECs because 
the density profile of the BEC is observable and thus plays an important role in observing  Kolmogorov turbulence. 
In this turbulence, Bogoliubov waves are significant, so  we call it Bogoliubov-wave turbulence in this paper.

Originally, in classical fluids, WT, which is turbulent flow dominated by waves,  has been studied \cite{wt1,wt2}. It is much different from HT because 
vortex structures are regarded as important in HT \cite{Davidson,Frisch}. 
There are weak and strong kinds of WT, depending on whether the nonlinearity is weak or strong.   
In weak WT, a constant-flux cascade of wave energy or wave action generates power-law behaviors in the spectrum, 
and Kolmogorov turbulence is known to occur in various wave systems such as those involving fluid surfaces (gravity and capillary waves) \cite{surface}, magnetic substances (spin waves) \cite{magnetic}, and elastic media (acoustic waves) \cite{elastic}. 

We expect that it is possible to confirm  Kolmogorov turbulence in  atomic BECs by the observation of the density profile in Bogoliubov-wave turbulence. 
In this paper, by applying  weak WT theory \cite{wt1,wt2} to the Gross-Pitaevskii (GP) equation, we theoretically and numerically study
the power-law behaviors for three spectra of a macroscopic wave function, density distribution, 
and Bogoliubov wave distribution, 
discussing the experimental possibility for observing  Kolmogorov turbulence by means of the density spectrum. 

There are some previous works on Bogoliubov-wave turbulence in  three-dimensional systems \cite{Dyachenko92,Zakharov05,Proment09}. 
In previous studies \cite{Dyachenko92,Zakharov05}  the $-3/2$ power law in the Bogoliubov-wave energy spectrum was analytically derived, and the equation of the fluctuation was derived with the assumption that the condensate function has a constant amplitude with a phase rotation induced by the chemical potential. Subsequent work \cite{Proment09} has suggested the $-3/2$ power law in the spectrum for the macroscopic wave function 
and a result consistent with this power law was numerically obtained.  

We reconsider this Bogoliubov-wave turbulence, analytically deriving the $-7/2$ power law in the spectrum for the macroscopic wave function and numerically confirming this power law. We find that the condensate dynamics induced by the fluctuation, which is neglected in the previous studies, is important for Bogoliubov-wave turbulence.
Furthermore, we focus on the spectrum for the density distribution, obtaining the $-3/2$ power law, and discuss the experimental possibility of this power law. 

The article is organized as follows. Section II describes the GP equation and the spectra for some 
quantities. In Sec. III, we apply weak WT theory to the GP equation, deriving the power laws for the spectra. In Sec. IV, we show our numerical result for this Bogoliubov-wave turbulence. Section V discusses comparison between the previous results and ours, the observation of Kolmogorov turbulence, and an inconsistency of our analytical calculation. Finally, we summarize our study in Sec. VI.

\section{Formulation}
We address a one-component BEC in a uniform system at zero temperature. This system is well described by the macroscopic wave function $\psi$
obeying the GP equation \cite{PS} given by
\begin{eqnarray}
i\hbar \frac{\partial}{\partial t} \psi  =  -\frac{\hbar^2}{2m} \nabla^2 \psi + g |\psi|^2 \psi  \label{GP}
\end{eqnarray}
with  particle mass $m$ and  interaction coefficient $g$. 

In this paper, we focus on spectra for the macroscopic wave function $\psi$, the density distribution $\rho = |\psi|^{2}$, and the Bogoliubov-wave distribution $b$.
The details of the Bogoliubov-wave distribution are defined in Sec. III B.
The spectrum for the wave function is defined by 
\begin{eqnarray}
C_{\rm w} (k) = \frac{1}{\triangle k} \sum _{k-\triangle k/2 \leq |\bm{k}_1| < k+\triangle k/2}  \langle |\bar{\psi} (\bm{k}_1) |^{2}  \rangle  \label{wave_correlation}
\end{eqnarray}
with a resolution of  $\triangle k = 2 \pi /L$ in  wave-number space and a system size $L$. 
The brackets indicate an ensemble average.
The function $\bar{\psi} (\bm{k})$ is the Fourier component of the macroscopic wave function calculated by 
$\mathcal{F} [ \psi (\bm{r})]$ with $\mathcal{F} [ \cdot ] = \int \cdot \hspace{1mm}  e^{-i\bm{k} \cdot \bm{r}} dV/L^d$ and  spatial dimension $d$. 
In the same way, we can define the spectra for the density and Bogoliubov-wave distributions as
\begin{eqnarray}
C_{\rm d} (k) = \frac{1}{\triangle k} \sum _{k-\triangle k/2 \leq |\bm{k}_1| < k+\triangle k/2}  \langle |\bar{\rho} (\bm{k}) |^{2} \rangle  ,  \label{density_correlation}
\end{eqnarray}
\begin{eqnarray}
C_{\rm b} (k) =\frac{1}{\triangle k}  \sum _{k-\triangle k/2 \leq |\bm{k}_1| < k+\triangle k/2}  \langle | b (\bm{k}) |^{2} \rangle  \label{bogo_correlation}
\end{eqnarray}
with the Fourier components of the density distribution $\bar{\rho}(\bm{k}) = \mathcal{F}[\rho(\bm{r})]$ and 
the Bogoliubov-wave distribution $b(\bm{k})$.
The physical meaning of these spectra is that $C_{\rm w}$  and $C_{\rm d}$ are correlation functions for the macroscopic wave function and density profile, and 
$C_{\rm b}$ is the Bogoliubov-wave distribution in the wave-number space.

\section{Application of weak wave turbulence theory}
We study  weak WT in a uniform BEC with the strong condensate, deriving the power exponents of  
$C_{\rm w} (k)$,  $C_{\rm d} (k),$ and $C_{\rm b} (k)$.
For this purpose, we apply weak WT theory to the GP equation (\ref{GP}). 
In this application, there is an important points, which is the nonlinear dynamics for the condensate. 
We take the squared terms of the fluctuation for the macroscopic wave function, treating this nonlinear dynamics. 
In this case, we can also use weak WT theory, obtaining these power exponents. 

\subsection{Equations of the condensate and the fluctuation}
We now derive the equations of  the condensate and the fluctuation. 
First, we define the condensate $\psi _0$ and the fluctuation $\phi$ as 
\begin{eqnarray}
\psi = \psi_0 ( 1+\phi ),   \label{fluctuation}
\end{eqnarray}
where $\psi_0$ is defined by 
\begin{eqnarray}
\psi_{0} = \frac{1}{L^d}\int \psi \hspace{1mm} dV .  \label{condensate}
\end{eqnarray}
From Eqs. (\ref{fluctuation}) and (\ref{condensate}), we can derive
\begin{eqnarray}
 \int \phi \hspace{1mm} dV = 0.   \label{fluctuation1}
\end{eqnarray}
Equations (\ref{condensate}) and (\ref{fluctuation1}) mean that the condensate and the fluctuation are the $\bm{k}=0$ and $\bm{k} \neq 0$ Fourier components of $\bar{\psi} (\bm{k})$, respectively.
In this paper, we consider a weak fluctuation from a strong condensate, assuming the weak nonlinear condition $| \phi| \ll  1$.

We substitute Eq. (\ref{fluctuation}) into the GP equation (\ref{GP}), deriving equations for $\psi_0$ and $\bar{\phi} (\bm{k}) = \mathcal{F}[\phi]$ within second order of the fluctuation:
\begin{eqnarray}
i\hbar \frac{\partial}{\partial t} \psi _0 =  g \rho_0 \psi_0 \Big[  1 +  \sum_{\bm{k}_1} \big(  2|\bar{\phi} (\bm{k}_1)|^2 + \bar{\phi} (\bm{k}_1)\bar{\phi} (-\bm{k}_1) \big)  \Big] ,  \label{GP_condensate}
\end{eqnarray}
\begin{eqnarray}
i\hbar \psi_0 \frac{\partial}{\partial t} \bar{\phi}(\bm{k}) = - i\hbar \bar{\phi}(\bm{k}) \frac{\partial}{\partial t} \psi_0 + \frac{\hbar^2 k^2} {2m} \psi_0 \bar{\phi}(\bm{k}) + g \rho_0 \psi_0 \Big[  2\bar{\phi}(\bm{k}) \nonumber \\
 + \bar{\phi}^{*}(-\bm{k})  +  2\sum_{\bm{k}_1 \bm{k}_2} \bar{\phi}^{*}(\bm{k}_1)\bar{\phi}(\bm{k}_2)\delta(\bm{k}+\bm{k}_1-\bm{k}_2) \nonumber \\
+  \sum_{\bm{k}_2 \bm{k}_3} \bar{\phi}(\bm{k}_2)\bar{\phi}(\bm{k}_3)\delta(\bm{k}-\bm{k}_2-\bm{k}_3)    \Big] ,  \label{GP_fluctuation1}
\end{eqnarray}
where $\rho _0$ is the condensate density $|\psi_0|^2$ and $\delta(\cdot)$ is the Kronecker delta.
Substituting Eq. (\ref{GP_condensate}) into Eq. (\ref{GP_fluctuation1}), we obtain the following equation for the fluctuation:
\begin{eqnarray}
i\hbar \frac{\partial}{\partial t} \bar{\phi}(\bm{k})  =  \frac{\hbar^2 k^2} {2m} \bar{\phi}(\bm{k}) + g \rho_0 \Big[  \bar{\phi}(\bm{k}) + \bar{\phi}^{*}(-\bm{k}) \nonumber \\  
+  2\sum_{\bm{k}_1 \bm{k}_2} \bar{\phi}^{*}(\bm{k}_1)\bar{\phi}(\bm{k}_2)\delta(\bm{k}+\bm{k}_1-\bm{k}_2) \nonumber \\
+  \sum_{\bm{k}_2 \bm{k}_3} \bar{\phi}(\bm{k}_2)\bar{\phi}(\bm{k}_3)\delta(\bm{k}-\bm{k}_2-\bm{k}_3)    \Big].  \label{GP_fluctuation2}
\end{eqnarray}
Therefore, we obtain Eqs. (\ref{GP_condensate}) and (\ref{GP_fluctuation2}) for the condensate and the fluctuation dynamics within the weak nonlinear condition. 

For  application of  weak WT theory, we rewrite Eq. (\ref{GP_fluctuation2}) into the canonical form given by 
\begin{eqnarray}
i\hbar \frac{\partial}{\partial t} \bar{\phi}(\bm{k})  =  \frac{\partial H}{\partial \bar{\phi}^{*}(\bm{k})},  \label{GP_fluctuation3}
\end{eqnarray} 
\begin{eqnarray}
H = H_2 + H_3,  \label{Hamiltonian2}
\end{eqnarray}
\begin{eqnarray}
H_2 &=&  \sum _{\bm{k}_{1}} \Biggl[ \Big( \frac{\hbar^{2}k_{1}^{2}}{2m} + g \rho_{0} \Big) |\bar{\phi}(\bm{k}_{1})|^{2} \nonumber \\ 
&+& \frac{g \rho_{0}}{2} \Big( \bar{\phi}(\bm{k}_{1}) \bar{\phi}(-\bm{k}_{1}) + \bar{\phi} ^{*}(\bm{k}_{1}) \bar{\phi} ^{*}(-\bm{k}_{1})  \Big) \Biggl]  \label{Hamiltonian2}
,\end{eqnarray}
\begin{eqnarray}
H_3 = g \rho_{0} \sum _{\bm{k}_1, \bm{k}_2, \bm{k}_3} \delta(\bm{k}_1 - \bm{k}_2 - \bm{k}_3) \nonumber \\ 
 \times  \Big( \bar{\phi}^{*}(\bm{k}_{1}) \bar{\phi}(\bm{k}_{2}) \bar{\phi}(\bm{k}_{3}) + \bar{\phi} (\bm{k}_{1}) \bar{\phi}^{*}(\bm{k}_{2}) \bar{\phi}^{*}(\bm{k}_{3})   \Big).   \label{Hamiltonian3}
\end{eqnarray}

Here, note that the condensate function $\psi_0$ has a time dependence. In  previous works \cite{Dyachenko92,Zakharov05,Proment09}, the condensate function $\psi_0$ was assumed to be $\sqrt{\bar{\rho} _{0} } {\rm exp}(-i\mu t/ \hbar)$ with  condensate density $\bar{\rho} _{0}= N_0/L^d$, the $\bm{k}=0$ particle number $N_{0}$ at the initial state, and the chemical potential $\mu = g \bar{\rho}_{0}$. 
However, we keep the second-order fluctuation terms in Eq. (\ref{GP_condensate}), so that the time dependence of $\psi_0$ is rather complicated. 
This term is very important for calculating the Bogoliubov-wave distribution, which is discussed in Sec. IV B.

\subsection{Diagonalization of the Hamitonian and Bogoliubov-wave distribution equation}
To diagonalize the one-body Hamiltonian $H_2$, we use the Bogoliubov transformation \cite{wt1} defined by 
\begin{eqnarray}
\bar{\phi}(\bm{k}) = u(k)b(\bm{k}) + v(k)b^{*}(-\bm{k}),  \label{Bogo_trans1}
\end{eqnarray}
\begin{eqnarray}
u(k) = \sqrt{ \frac{1}{2} \Big( \frac{\epsilon _{0}(k) + g \rho_{0}}{\epsilon_{\rm b}(k)} +1 \Big)  },   \label{Bogo_trans2}
\end{eqnarray}
\begin{eqnarray} 
v(k) = - \sqrt{ \frac{1}{2} \Big( \frac{\epsilon _{0}(k) + g \rho_{0}}{\epsilon_{\rm b}(k)} -1 \Big) },   \label{Bogo_trans3}
\end{eqnarray}
where $b(\bm{k})$ is the canonical variable for  Bogoliubov waves (Bogoliubov-wave distribution) and $\epsilon _{\rm b}(k)$ is the dispersion relation for this wave defined by 
\begin{eqnarray}
\epsilon _{\rm b}(k) = \sqrt{\epsilon _{0}(k)(\epsilon _{0}(k) + 2g \rho_{0})}  \label{Bog_dis}
\end{eqnarray}
with $\epsilon _{0}(k) = \hbar^{2} k^{2} /2m$. Applying this transformation to the Hamiltonian of Eqs. (\ref{Hamiltonian2}) and (\ref{Hamiltonian3}) leads to
\begin{eqnarray}
H_2 = \sum _{\bm{k}_{1}} \epsilon _{\rm b}(k_1) |b(\bm{k}_{1})|^{2},  \label{Hamiltonian4}
\end{eqnarray}
\begin{eqnarray}
H_3= \sum _{\bm{k}_1, \bm{k}_2, \bm{k}_3} \delta(\bm{k}_1 - \bm{k}_2 - \bm{k}_3)  V(\bm{k}_1, \bm{k}_2, \bm{k}_3)  \nonumber \\ 
\times \Big( b^{*}(\bm{k}_1)b(\bm{k}_2)b(\bm{k}_3) + b(\bm{k}_1) b^{*}(\bm{k}_2)b^{*}(\bm{k}_3)  \Big) \nonumber \\
+ \sum _{\bm{k}_1, \bm{k}_2, \bm{k}_3} \delta(\bm{k}_1 + \bm{k}_2 + \bm{k}_3)  W(\bm{k}_1, \bm{k}_2, \bm{k}_3)  \nonumber \\ 
\times \Big( b^{*}(\bm{k}_1)b^{*}(\bm{k}_2)b^{*}(\bm{k}_3) + b(\bm{k}_1)b(\bm{k}_2)b(\bm{k}_3) \Big),    \label{Hamiltonian5}
\end{eqnarray}
where $V$ and $W$ are the interaction functions for  Bogoliubov waves.  These functions are given by 
\begin{eqnarray}
V(\bm{k}_1, \bm{k}_2, \bm{k}_3) = g \rho_{0} \Big( u_1 u_2 u_3 + v_1 v_2 u_3 \nonumber \\ 
 + v_1 u_2 v_3 + v_1 v_2 v_3 + u_1 v_2 u_3 + u_1 u_2 v_3 \Big),  \label{Interaction1}
\end{eqnarray}
\begin{eqnarray}
W(\bm{k}_1, \bm{k}_2, \bm{k}_3) = g \rho_{0} \Big( u_1 v_2 v_3 + v_1 u_2 u_3 \Big),  \label{Interaction2}
\end{eqnarray}
where we use  simple expressions such as $u_1$ and $v_1$ for $u(k_1)$ and $v(k_1)$.

We must notice that the Bogoliubov coefficients $u(k)$ and $v(k)$ have time dependence through the condensate density $\rho_0$.  
Thus, the transformation of Eq. (\ref{Bogo_trans1}) is not the canonical transformation. However, as shown in the following, the time dependence of these coefficients are 
found to give only small corrections in the weak nonlinear condition, so that the transformation of Eq. (\ref{Bogo_trans1}) approximately becomes the canonical 
transformation. To indicate this, we derive the time development equation for $\rho _0$ from Eq. (\ref{GP_condensate}), which is given by
\begin{eqnarray}
\frac{\partial}{\partial t} \rho_0 = \frac{g\rho_0^2}{i\hbar} \sum_{\bm{k}_1}  \Big[  b(\bm{k}_1)b(-\bm{k}_1)-b^*(\bm{k}_1)b^*(-\bm{k}_1)  \Big]    \label{density_equ}.
\end{eqnarray}
Then, the time derivative of the Bogoliubov coefficient $u(k)$ becomes 
\begin{eqnarray}
\frac{\partial}{\partial t} u(k) = \frac{\partial u(k)}{\partial \rho_0}  \frac{\partial \rho_0}{\partial t},   
\end{eqnarray}
which means that the value of this derivative is the second order of the fluctuation. The same thing is true in the time derivative of $v(k)$.   
As a result, in the weak nonlinear condition, the time derivative of Eq. (\ref{Bogo_trans1}) is
\begin{eqnarray}
\frac{\partial}{\partial t} \bar{\phi}(\bm{k}) \simeq u(k) \frac{\partial}{\partial t} b(\bm{k}) + v(k) \frac{\partial}{\partial t}b^{*}(-\bm{k})
\end{eqnarray}
since the third order of the fluctuation is neglected in Eq. (\ref{GP_fluctuation2}).
Thus, the Bogoliubov transformation of Eq. (\ref{Bogo_trans1}) approximately becomes the canonical transformation. 

Based on the above calculation, the time development equation for $b(\bm{k})$ becomes 
\begin{eqnarray}
i\hbar \frac{\partial}{\partial t} b(\bm{k})  =  \frac{\partial H}{\partial b^{*}(\bm{k})}.  \label{Bogo_equ}
\end{eqnarray}
This equation can describe weak nonlinear dynamics for  Bogoliubov waves. 
In the following subsection, we apply  weak WT theory to Eqs. (\ref{density_equ}) and (\ref{Bogo_equ}), discussing the behavior of the spectra 
for the wave function, density distribution, and Bogoliubov-wave distribution. 

\subsection{Kinetic equation for  Bogoliubov waves}
In weak WT theory, the kinetic equation for the spectrum of the canonical variable $b(\bm{k})$ can be derived \cite{wt1,wt2}. 
The spectrum is defined by 
\begin{eqnarray}
n(\bm{k}) = \Big( \frac{L}{2 \pi} \Big)^d  \langle |b(\bm{k})|^2 \rangle.   \label{Bogo_cor}
\end{eqnarray} 
Applying  weak WT theory to Eqs. (\ref{density_equ}) and (\ref{Bogo_equ}), we obtain the following kinetic equation:
\begin{eqnarray}
\frac{\partial}{\partial t} n(\bm{k}) = \int \Big( \mathcal{R}(\bm{k},\bm{k}_1,\bm{k}_2) - \mathcal{R}(\bm{k}_1,\bm{k}_2,\bm{k}) \nonumber \\
 - \mathcal{R}(\bm{k}_2,\bm{k},\bm{k}_1) \Big) d\bm{k}_1 d\bm{k}_2,   \label{kin_equ1}
\end{eqnarray}
\begin{eqnarray}
\mathcal{R}(\bm{k},\bm{k}_1,\bm{k}_2) = 2 \pi |V(\bm{k},\bm{k}_1,\bm{k}_2)|^2  \nonumber \\
\times \delta _{\rm d}(\epsilon_{\rm b} (k)-\epsilon_{\rm b} (k_1)-\epsilon_{\rm b}(k_2)) \delta _{\rm d} (\bm{k}-\bm{k}_1-\bm{k}_2) \nonumber \\ 
\times ( n(\bm{k}_1)n(\bm{k}_2) - n(\bm{k}_2)n(\bm{k})- n(\bm{k})n(\bm{k}_1) )   \label{kin_equ2}
\end{eqnarray}
with the Dirac delta function $\delta _{\rm d}(\cdot)$.
In this application, the contributions from $b_1 b_2 b_3$ and $b_1^* b_2^* b_3^*$ terms in Eq. (\ref{Hamiltonian5}) are found to be smaller than those of other terms \cite{wt2,Zakharov05,Lvov97} because these terms accompany $\delta _{\rm d} (\epsilon_{\rm b} (k)+\epsilon_{\rm b} (k_1)+\epsilon_{\rm b}(k_2))$. In  Bogoliubov waves, the argument in this Dirac delta function is never satisfied, so that  the interaction $W(\bm{k}_1, \bm{k}_2, \bm{k}_3)$ does not appear in Eqs. (\ref{kin_equ1}) and (\ref{kin_equ2}). The details of this calculation are described in \cite{wt2}, where a nonlinear canonical transformation causes the interaction $W(\bm{k}_1, \bm{k}_2, \bm{k}_3)$ to vanish. 
The transformation does not change the expression for $V(\bm{k}_1, \bm{k}_2, \bm{k}_3)$, but it does change the expression for $b(\bm{k})$. However, this change gives only a small contribution to $n(\bm{k})$.

Note that the application of weak WT theory to Eqs. (\ref{density_equ}) and (\ref{Bogo_equ}) is slightly different from usual cases where the dispersion relation $\epsilon _{\rm b}$ and the interaction $V$ do not have time dependence. In our case, the condensate density $\rho_0$ depends on time, which makes $\epsilon _{\rm b}$ and $V$ dependent on time. Thus, we cannot use the conventional weak WT theory in \cite{wt2}, but we can derive the kinetic equation (\ref{kin_equ1}). The detail of this derivation is described in Appendix A. 

We are interested in the large-scale dynamics of  Bogoliubov waves because, in experiments, it may be difficult to observe 
small-scale dynamics. Here, to elucidate the meaning of large scale, let us note the dispersion relation for  Bogoliubov waves.
From Eq. (\ref{Bog_dis}), it follows that the $k$ dependence of the dispersion relation drastically changes depending on whether or not the amplitude of the wave number is larger than $k_{\rm b} = 2 \sqrt{mg\rho _0}/\hbar$. In the wave-number region smaller than $k_{\rm b}$, the dispersion relation becomes phononlike, which is expressed by $\hbar c_{\rm s} k$ with  sound velocity $c_{\rm s} = \sqrt{g \rho_{0}/m}$, while in the region larger than $k_{\rm b}$ it behaves like that of a free particle. From this property of the dispersion relation, in this paper, the large (small) scale means the wave-number region is smaller (larger) than $k_{\rm b}$.

Finally, we must comment on the importance of the dimension of the kinetic equation in the low-wave-number region where $\epsilon _{\rm b}(k)$ has  linear dispersion. 
In this region,  Bogoliubov-wave turbulence is similar to acoustic turbulence, for which, in a two-dimensional system, the validity of the kinetic equation is controversial \cite{wt1,Fal,Lvov97}. Thus, in the following, we address three-dimensional systems.

In the next section, to investigate the behavior of the spectra in the low-wave-number region, 
we calculate an expression of the interaction $V$ for the low-wave-number limit. 

\subsection{Interaction of  Bogoliubov waves in the low-wave-number region}
We now derive the expression for $V(\bm{k}_1, \bm{k}_2, \bm{k}_3)$ in the low-wave-number region. 
First, we assume that this interaction is local in the turbulent state, which means that the dominant contribution 
to the interaction comes from wave numbers of the same order \cite{Zakharov05}. 
This locality is also discussed in Appendix D. 
Thus, the wave-number region $k_{j} < k_{\rm b}$ ($j=1,2,3$) is considered in the following. 

In the low-wave-number region, the Bogoliubov coefficients $u(k)$ and $v(k)$ are approximated as
\begin{eqnarray}
u(k) \simeq \frac{1}{2}\sqrt{\frac{k_{\rm b}}{k}} \Biggl[ 1 + \frac{k}{k_{\rm b}} + \frac{1}{4} \Bigl( \frac{k}{k_{\rm b}} \Bigl)^{2} - \frac{1}{4} \Bigl( \frac{k}{k_{\rm b}} \Bigl)^{3}    \Biggl],  \label{Bogo_coe1}
\end{eqnarray}
\begin{eqnarray}
v(k) \simeq  \frac{1}{2}\sqrt{\frac{k_{\rm b}}{k}} \Biggl[ -1 + \frac{k}{k_{\rm b}} - \frac{1}{4} \Bigl( \frac{k}{k_{\rm b}} \Bigl)^{2} - \frac{1}{4} \Bigl( \frac{k}{k_{\rm b}} \Bigl)^{3}    \Biggl].  \label{Bogo_coe2}
\end{eqnarray}
We substitute Eqs. (\ref{Bogo_coe1}) and (\ref{Bogo_coe2}) into the interaction $V$ of Eq. (\ref{Interaction1}), obtaining the following expression:
\begin{eqnarray}
V(\bm{k}_1, \bm{k}_2, \bm{k}_3) \simeq \frac{g \rho_{0}}{8} \sqrt{\frac{k_{\rm b}^{3}}{k_1 k_2 k_3}} \Biggl[ -\frac{2}{k_{\rm b}} ( k_1 - k_2 - k_3  ) \nonumber \\
+ \frac{1}{2k_{\rm b}^{3}} ( k_1^3 - k_2^3 - k_3^3 + k_1^2 k_2 -  k_1 k_2^2 \nonumber \\ 
 +  k_2^2 k_3 +  k_2 k_3^2 +  k_1^2 k_3 -  k_1 k_3 ^2 + 12  k_1 k_2 k_3)    \Biggl].    \label{Interaction3}
\end{eqnarray}

As shown by the two delta functions of Eq. (\ref{kin_equ2}), Bogoliubov waves must satisfy both momentum- and energy-conservation laws, which are expressed by 
$\bm{k}_{1} = \bm{k}_{2} + \bm{k}_{3} $ and $ \epsilon _{\rm b}(k_1) =  \epsilon _{\rm b}(k_2) +  \epsilon _{\rm b}(k_3) $. 
These conservations lead to the rather simple form of Eq. (\ref{Interaction3}). 
In the low-wave-number region, the dispersion relation becomes 
\begin{eqnarray}
\epsilon_{\rm b} (k) \simeq \hbar c_{\rm s} k \Big[1+ \frac{1}{2}\Big(\frac{k}{k_{\rm b}} \Big)^2 \Big]. \label{disper}
\end{eqnarray}
As a result, the effective interaction $V_{\rm b}$ for  Bogoliubov waves becomes
\begin{eqnarray}
V_{\rm  b}(\bm{k}_1, \bm{k}_2, \bm{k}_3) = \frac{3g\rho_0}{2 k_{\rm b}^{3/2}} \sqrt{k_1k_2k_3}.    \label{Interaction4}
\end{eqnarray}
The derivation of this interaction is described in Appendix B. 
This interaction satisfies the scaling law $V_{\rm  b}( \lambda \bm{k}_1, \lambda \bm{k}_2, \lambda \bm{k}_3) = \lambda^{3/2} V_{\rm  b}(\bm{k}_1, \bm{k}_2, \bm{k}_3)$, which is directly related to the power exponent of the spectra.
In previous studies different canonical variables were used to  derive the interaction function, which is different from Eq. (\ref{Interaction4}), but the scaling law is the same \cite{Dyachenko92,Zakharov05}. 

\subsection{Derivation of power exponents of the spectra}
In a stationary state, we can derive the power exponents of the spectra by using the kinetic equation (\ref{kin_equ1}) with 
the interaction of Eq. (\ref{Interaction4}). Our derivation makes use of the fact that a Bogoliubov-wave energy flux is constant in the wave-number space in the weak WT. 

We integrate the kinetic equation (\ref{kin_equ1}) with the effective interaction (\ref{Interaction4}) and the dispersion relation (\ref{disper}), 
obtaining  
\begin{eqnarray}
\frac{\partial}{\partial t} n(k) = I (k), \label{bogkin1}
\end{eqnarray}
\begin{eqnarray}
I (k) = \frac{9\pi^2 g^2 \rho_0^2}{\hbar c_{\rm s}k_{\rm b}^3} \int_0^{\infty} \Big[ k_1^2 (k-k_1)^2 f_1(k,k_1) \nonumber \\
- k_1^2(k_1-k)^2 f_1(k_1,k)  \nonumber \\
- k_1^2(k+k_1)^2 f_2(k,k_1) \Big] dk_1,  \label{bogkin2}
\end{eqnarray}
\begin{eqnarray}
f_1(k,k_1) = \theta_{\rm H}(k-k_1)\Big( n(k_1)n(k-k_1)-n(k)n(k_1) \nonumber \\
-n(k)n(k-k_1) \Big),  \label{bogkin3}
\end{eqnarray}
\begin{eqnarray}
f_2(k,k_1) = n(k_1)n(k)-n(k+k_1)n(k)  \nonumber \\ 
-n(k_1)n(k+k_1) \label{bogkin4}
\end{eqnarray}
with the Heaviside step function $\theta _{\rm H} (\cdot)$. 
Here we assume the isotropy of $n(\bm{k})=n(k)$, which is reasonable in isotropic turbulence. 
In Appendix C, we show the derivation of Eqs. (\ref{bogkin1}) - (\ref{bogkin4}).

From Eq. (\ref{bogkin1}), we can derive the continuity equation of the Bogoliubov-wave energy spectrum given by 
\begin{eqnarray}
\frac{\partial}{\partial t}\mathcal{E}_{\rm b}(k) + \frac{\partial}{\partial k} \Pi (k) = 0, \label{flux1}
\end{eqnarray}
where $\mathcal{E}_{\rm b}(k)$ and $\Pi (k)$ are the Bogoliubov-wave energy spectrum and the energy flux defined by 
\begin{eqnarray}
\mathcal{E}_{\rm b}(k) &=&  \int  \epsilon _{\rm b}(k) n(k) d\Omega_{k}  \nonumber \\
&=& 4 \pi k^2  \epsilon _{\rm b}(k) n(k), 
\end{eqnarray}
\begin{eqnarray}
\Pi (k) = -4 \pi \int _{0}^{k} k_1^{2}  \Big[  \epsilon _{\rm b}(k_1) I(k_1) + n(k_1) \frac{\partial}{\partial t} \epsilon_{\rm b}(k_1)  \Big] dk_1.  \label{flux2}
\end{eqnarray}

In the statistically stationary state, the time derivatives of $\mathcal{E}_{\rm b}(k)$ in Eq. (\ref{flux1}) is zero, which means that 
the energy flux $\Pi (k)$ is independent of the wave number. Also, the time derivative of $\epsilon_{\rm b}(k)$ is zero because the condensate density $\rho_0$ is stationary in this state. Assuming the power law $n(k) \propto k^{-m}$ and applying the transformation $k_1=k \tilde{k}$ to 
Eq. (\ref{flux2}), we obtain 
 \begin{eqnarray}
\Pi (k) \propto k^{9-2m},   \label{flux3}
\end{eqnarray}
which exhibits that the energy flux is constant if the power exponent $m$ is $9/2$. 
Thus, we derive the power law in the Bogoliubov-wave turbulence: 
\begin{eqnarray}
n(k) \propto k^{-9/2}.   \label{b_sol1}
\end{eqnarray}
This power law is found to be the local, so that the locality assumption in Sec. IV D is justified, which is described in Appendix D.  
Therefore, the spectrum integrated over the solid angle yields
\begin{eqnarray}
C_{\rm b}(k) \propto k^{-5/2}.   \label{b_sol2}
\end{eqnarray}

We can derive the power exponent for $C_{\rm w}$ by using Eq. (\ref{b_sol2}). In the low-wave-number region, the relation between $\bar{\phi} (\bm{k})$ and $b(\bm{k})$ becomes
\begin{eqnarray}
\bar{\phi}(\bm{k}) \propto \frac{b(\bm{k})}{\sqrt{k}} - \frac{b^{*}(-\bm{k})}{\sqrt{k}},   \label{b_w_rel1}
\end{eqnarray}
which is obtained from Eq. (\ref{Bogo_trans1}).
Then, we assume  isotropy for $b(\bm{k})$ and use the approximation to derive Eq. (\ref{kin_equ1}), obtaining
\begin{eqnarray}
\langle |\bar{\phi}(k)|^2 \rangle \propto \frac{1}{k} \langle |b(k)|^2 \rangle .
\label{b_w_rel2}
\end{eqnarray}
As a result, in the low-wave-number region, the spectrum for the macroscopic wave function exhibits the power law behavior given by
\begin{eqnarray}
C_{\rm w}(k) & \propto & k^{-1} C_{\rm b}(k)  \nonumber \\ 
& \propto & k^{-7/2}. \label{b_w_rel3}
\end{eqnarray}
In this derivation, we use the Fourier component $\mathcal{F}[\psi] = \psi_0( \delta (\bm{k}) + \bar{\phi}(\bm{k}))$.

For the density spectrum, the following expression for $\psi$ is useful to derive the power exponent of $C_{\rm d}$:
\begin{eqnarray}
\psi(\bm{r}) &=& \sqrt{\rho_0 +\delta \rho(\bm{r})} \hspace{1mm} {e}^{i (\theta_0 + \delta \theta(\bm{r}))} \nonumber \\
 &\simeq& \psi_0\left( 1 + \frac{\delta \rho(\bm{r})}{2\rho_0} + i \delta \theta(\bm{r})\right) ,  \label{w_d_p_rel1}
\end{eqnarray}
where $\psi_0$ is written as $\sqrt{\rho_0} {\rm exp}(i \theta _0) $, and $\delta \rho$ and $\delta \theta$ are the density and phase fluctuations, respectively, around it.
Then, the fluctuations of the wave function and the density are obtained from
\begin{eqnarray}
\phi (\bm{r})= \frac{\delta \rho(\bm{r})}{2 \rho_0} + i \hspace{1mm} \delta \theta (\bm{r}),  \label{w_d_p_rel2}
\end{eqnarray}
\begin{eqnarray}
\delta \rho (\bm{r}) &=& \rho_0 \Big( \phi (\bm{r}) +  \phi ^{*} (\bm{r}) \Big),   \label{w_d_p_rel3}
\end{eqnarray}
from which we can express the Fourier component of $\delta \rho$ with the canonical variable $b$.
Performing a similar calculation for Eqs. (\ref{b_w_rel1})--(\ref{b_w_rel3}), we can derive the power law in the spectrum for the density distribution:
\begin{eqnarray}
C_{\rm d}(k)  &\propto& kC_{\rm b}(k) \nonumber \\
 &\propto&  k^{-3/2}.  \label{w_d_p_rel4}
\end{eqnarray}
We consider that this power law is a candidate for an experimentally observable quantity to confirm the presence of Kolmogorov turbulence.

As shown in above, the power exponents of $C_{\rm w}$ and $C_{\rm d}$ are related to $C_{\rm b}$ through the behavior of Bogoliubov coefficients of 
Eqs. (\ref{Bogo_trans2}) and (\ref{Bogo_trans3}) in the low wave number region. 
Thus, the difference of the power exponents reflects the property of the collective mode.   

In the next section, by numerically calculating the GP equation, we discuss these power laws. 

\section{Numerical results} 
We now present our numerical results for Bogoliubov-wave turbulence with the GP equation. 
One of the main results is in good agreement with the power exponents of Eqs. (\ref{b_sol2}), (\ref{b_w_rel3}), and (\ref{w_d_p_rel4}) 
derived by using weak WT theory.  

\begin{figure} [b]
\begin{center}
\includegraphics[width=85mm]{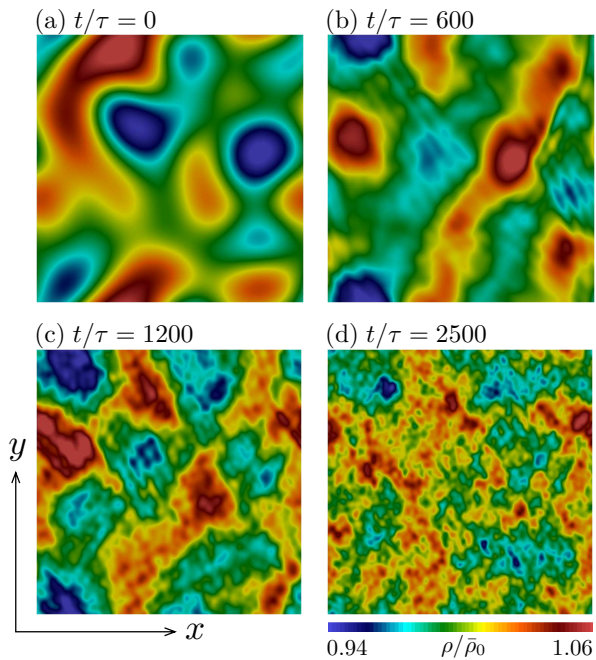}
\caption{(Color online) Spatial distribution of the density $\rho(x,y,z=128\xi)$ at $t/\tau=$ (a) 0, (b) 600, (c) 1200, and (d) 2500.
The smaller structures are nucleated in turn as  time passes, providing evidence of the direct energy cascade. }
\label{fig1}
\end{center}
\end{figure}

\begin{figure} [t]
\begin{center}
\includegraphics[width=90mm]{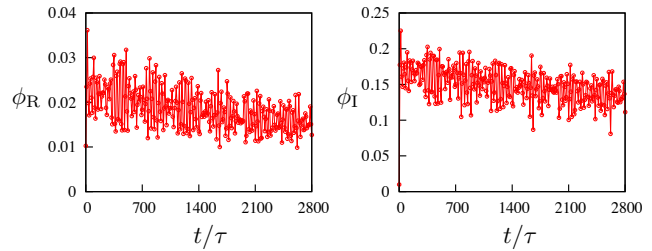}
\caption{(Color online) Time development of the spatially averaged absolute values $\phi _{\rm R}$ and $\phi _{\rm I}$ for real and imaginary parts of the fluctuation $\phi$. These graphs exhibit a system satisfying the weak nonlinear condition. }
\label{fig2}
\end{center}
\end{figure}

\begin{figure} [t]
\begin{center}
\includegraphics[width=90mm]{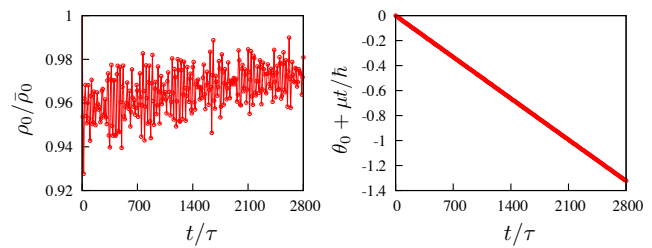}
\caption{(Color online)  Time development of the amplitude $\rho _0$ and phase $\theta _0$ of the condensate $\psi_0$. 
The macroscopic wave function is found not to obey $\sqrt{\bar{\rho} _{0}} {\rm exp}(-i\mu t/ \hbar)$. 
Particularly, the deviation of the phase is large. }
\label{fig3}
\end{center}
\end{figure}

\subsection{Numerical method}
Our numerical calculation treats a three-dimensional BEC without a trapping potential. 
Time and length are normalized by $\tau = \hbar/g \bar{\rho}_{\rm 0}$ and $\xi = \hbar/\sqrt{2mg \bar{\rho}_{\rm 0}}$, and the numerical system size $L \times L  \times L$ is $256\xi \times 256\xi \times 256\xi$ with spatial resolution $dx/\xi = 1$. Because of this resolution, the quantized vortex dynamics such as nucleation, reconnection, etc. cannot be described correctly since its core size is the order of $\xi$. However, we address the turbulence without the vortices, so that it is sufficient to use this resolution parameter. In this situation, we numerically solve the GP equation by using the pseudo-spectral method. The time propagation is performed by using a fourth-order Runge-Kutta method with  time resolution $dt /\tau = 5 \times 10^{-3}$.

We use an initial state in which energy is injected in the large scale  to confirm the direct energy cascade 
solutions corresponding to Eq. (\ref{b_sol2}), (\ref{b_w_rel3}), and (\ref{w_d_p_rel4}). 
The initial state is prepared by the random numbers as follows:
\begin{eqnarray}
\psi (\bm{r}) = \sqrt{\bar{\rho}_{\rm 0}} \Big(1 + \phi (\bm{r}) \Big),
\end{eqnarray}
\begin{eqnarray}
\mathcal{F}[\phi](\bm{k}) =  3000(R_1 + i R_2)k^4\xi^4 {\rm exp}(-k^2 \xi ^2/0.0016), 
\end{eqnarray}
where $R_1$ and $R_2$ are the random numbers in the range $[0.5,-0.5)$.
Figure 1(a) shows the density distribution in the initial state and demonstrates that the energy is injected at the large scale.

In our calculation, dissipation is phenomenologically included by replacing $i\hbar \partial / \partial t$ with 
$(i- \gamma \theta_{\rm H}(k-\bar{k}_{\rm b}))\hbar \partial / \partial t$ in the Fourier transformed Eq. (\ref{GP}) \cite{KT05,KT07}.
Here, $\bar{k}_{\rm b}$ is $2\sqrt{mg\bar{\rho}_0}/\hbar$, and  $\gamma$ is the strength of the dissipation. 
In our calculation, we use $\gamma = 0.03$. 
We comment on the dissipation region. 
In this paper, we focus on the low-wave-number region, adopting the dissipation working in the region larger than $\bar{k}_{\rm b}$. 
There is arbitrariness in the choice of the expression for the dissipation.
However, in a previous paper \cite{dissipation}, the dissipation was found to work in the high-wave-number region at low temperature, so that we use 
this expression for the dissipation.

In summary, we prepare the unstable initial state for the confirmation of the direct cascade, numerically calculating the GP equation with the phenomenological dissipation. In this calculation, we do not force the system, so that this is decaying turbulence. 

\subsection{Time development of the condensate and the fluctuation}
We numerically confirm that our turbulence satisfies the weak nonlinear condition $|\phi| \ll 1$.  
Figure 2 shows the time dependence of the spatially averaged absolute values $\phi _{\rm R}$ and $\phi _{\rm I}$ for real and imaginary parts of the fluctuation $\phi$, 
from which the order of the fluctuation $|\phi|$ is found to be about $0.13$. 

Figure 3 shows the time development of the amplitude $\rho _0$ and phase $\theta _0$ of the condensate $\psi_0$; we find 
that the condensate does not obey $\sqrt{\bar{\rho} _{0}} {\rm exp}(-i\mu t/ \hbar)$. 
This means that the fluctuation term in Eq. (\ref{GP_condensate}) affects the dynamics of the condensate. 
This effect is very important for calculating the Bogoliubov-wave distribution $b(\bm{k})$, which is discussed in Sec. IV C.  

\begin{figure} [t]
\begin{center}
\includegraphics[width=81mm]{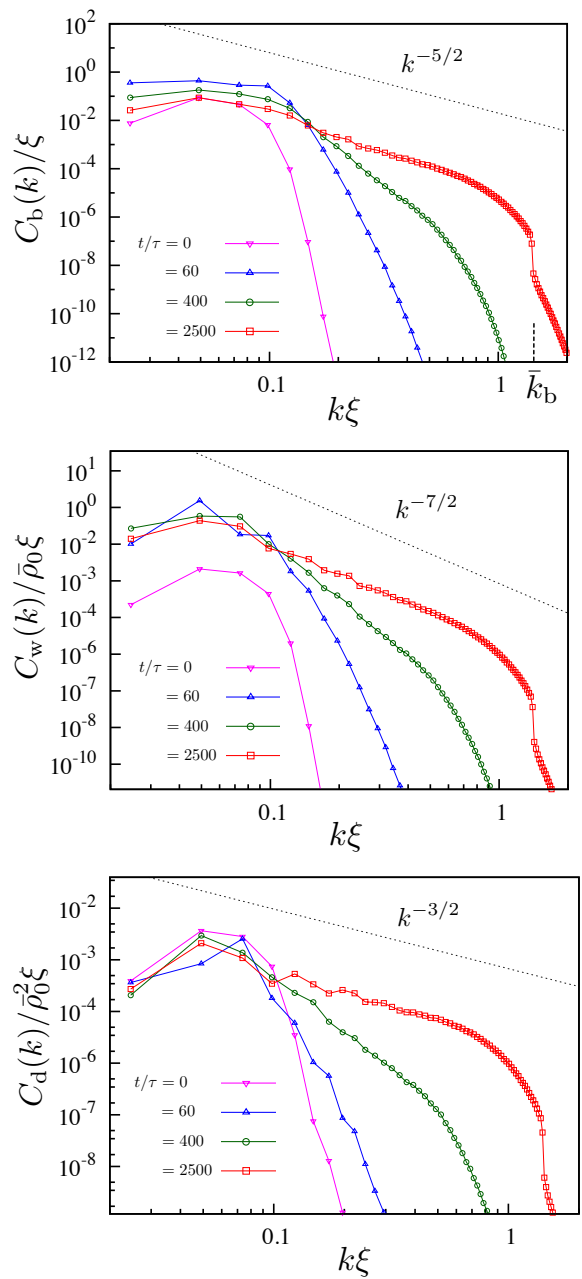}
\caption{(Color online) Time development of the spectra for the Bogoliubov-wave distribution $C_{\rm b}$, the macroscopic wave function $C_{\rm w}$, and the density distribution $C_{\rm d}$. The spectra are averaged over six calculations with different initial noise. 
The direct energy cascade occurs, and the power exponents of Eq. (\ref{b_sol2}), (\ref{b_w_rel3}), and (\ref{w_d_p_rel4}) show  good agreement with the numerical results. }
\label{fig4}
\end{center}
\end{figure}

\subsection{Time development of the spectra}
We next numerically calculate the spectra $C_{\rm b}$, $C_{\rm w}$, and $C_{\rm d}$, finding that 
the numerical results show good agreement with our analytical results of Eqs. (\ref{b_sol2}), (\ref{b_w_rel3}), and (\ref{w_d_p_rel4}).
Figure 4 shows the time development of the three spectra, demonstrating the direct energy cascade from small to large wave numbers. 
This cascade is also confirmed by the density distribution as shown in Figs. \ref{fig1}(a)--\ref{fig1}(d), in which small-scale structure is seen to grow. 
In each spectrum, as  time passes, the power-law behaviors corresponding to Eq. (\ref{b_sol2}), (\ref{b_w_rel3}), and (\ref{w_d_p_rel4}) gradually grow, and the scaling range finally becomes $0.05 \alt k\xi \alt 0.6$. The fluctuation in $k\xi \alt 0.2$ is large, which may be the finite-size effect. Particularly, the time development of $C_{\rm b}$ is large. 

We note the calculation of the Bogoliubov-wave distribution $b(\bm{k})$. As pointed out in Sec. IV B, in  previous works \cite{Dyachenko92,Zakharov05,Proment09}, 
the condensate function is assumed to be $\sqrt{\bar{\rho} _{0}} {\rm exp}(-i\mu t/ \hbar)$. When we use this condensate function and calculate the 
Bogoliubov-wave distribution, the spectrum $C_{\rm b}$ does not show a $-5/2$ power law and the deviation from this law is 
much larger. When we use the condensate function defined by Eq. (\ref{condensate}), however, the spectrum $C_{\rm b}$ shows 
good agreement with Eq. (\ref{b_sol2}). Therefore, it is important to take account of the dynamics of the condensate coupled with the fluctuation. 
Such a condensate has been called a quasicondensate \cite{Nazarenko14}, but the condensate dynamics 
of Eq. (\ref{GP_condensate}) has not been treated in  previous studies. 

Finally, we comment on the relation between the power exponents and the strength of the fluctuation. 
Due to our analytical calculation, if the weak nonlinear condition is satisfied, the energy flux is small, and the power exponents of Eqs. (\ref{b_sol2}), (\ref{b_w_rel3}), and (\ref{w_d_p_rel4}) can appear, although we do not numerically confirm the appearance of the same power laws 
in the case of weaker fluctuation because it takes a much longer time to do the numerical calculation. 
On the other hand, if the fluctuation is strong, the energy flux is large, and other power exponents may appear. 
Such a strong WT is the future work. 

\section{Discussion}
In this section, we derive the three power exponents for $C_{\rm b}$, $C_{\rm w}$, and $C_{\rm d}$. 
In Sec. V A, we  detail  the difference between the power exponents in previous  results and ours. 
In Sec. V B, we propose some conditions to observe Kolmogorov turbulence in atomic BECs and comment on the turbulence theory beyond the weak WT one. 
We consider that our result for $C_{\rm d}$ may be the key to achieve this observation.
In Sec. V C, we discuss an inconsistency of our analytical calculation in Sec. III. 

\subsection{Comparison between previous results and our results}
In a previous study \cite{Proment09}, the spectrum of the macroscopic wave function was suggested to obey a $-3/2$ power law in 
a three-dimensional system. However, our result has another $-7/2$ power law.  

To discuss the reason for this difference, we review the derivation of the $-3/2$ power law in this previous paper \cite{Proment09}. 
In  \cite{Zakharov05}, Zahkarov and Nazarenko studied  Bogoliubov-wave turbulence  using the two canonical variables 
$\varphi = \sqrt{\rho} $ and $\Pi=2 \hbar \rho \theta$ with $\psi = \sqrt{\rho} {\rm exp} (i\theta)$. 
To study the fluctuation around the strong condensate, they took expressions $\varphi  = \varphi _{0} + \varphi _{1}$ and $\Pi=\Pi_{1}$, where 
the quantity $\varphi _{0}$ corresponds to the condensate and $\varphi _{1}$ and  $\Pi_{1}$ correspond to the fluctuations. 
By using the following canonical transformation for these canonical variables, 
they introduced new canonical variables $B(\bm{k})$, diagonalizing a one-body Hamiltonian as 
\begin{eqnarray}
H_2' = \frac{1}{\hbar}\sum _{\bm{k}_1} \epsilon _{\rm b} (k) |B(\bm{k}_1)|^2. 
\end{eqnarray}
The canonical variables $B(\bm{k})$ were defined by 
\begin{eqnarray}
\bar{\varphi}_1(\bm{k}) =  \frac{1}{2}\sqrt{\frac{\epsilon_{0}(k)}{\hbar \epsilon_{\rm b}(k)}}  \Big( B(\bm{k}) + B^*(-\bm{k})  \Big),  
\end{eqnarray}
\begin{eqnarray}
\bar{\Pi}_1(\bm{k}) =  -i\sqrt{\frac{\hbar \epsilon_{\rm b}(k)}{\epsilon_{0}(k)}}  \Big( B(\bm{k}) - B^*(-\bm{k})  \Big), 
\end{eqnarray}
with the Fourier components $\bar{\varphi}_1(\bm{k}) =\mathcal{F}[\varphi_1]$ and $\bar{\Pi}_1(\bm{k}) =\mathcal{F}[\Pi_1]$.
Here, note that $H_2'$ is different from $H_2$ of Eq. (\ref{Hamiltonian4}). Then, they derived the $-3/2$ power law in the Bogoliubov-energy spectrum defined by 
\begin{eqnarray}
\mathcal{E}_{\rm B}(k) =  \frac{1}{\hbar \triangle k} \sum _{k-\triangle k/2 \leq |\bm{k}_1| < k+\triangle k/2} \epsilon _{\rm b}(k) \langle |B(\bm{k})|^2\rangle . 
\end{eqnarray}
The previous paper \cite{Proment09} makes use of this power law to derive the power exponent of the spectrum for the macroscopic wave function. 
In this previous paper $\mathcal{E}_{\rm B}(k)$ is estimated as  
\begin{eqnarray}
\frac{1}{\triangle k} \sum _{k-\triangle k/2 \leq |\bm{k}_1| < k+\triangle k/2} \epsilon _{\rm b}'(k) \langle |\bar{\psi}(\bm{k})|^2 \rangle, \label{previous}
\end{eqnarray}
with $\epsilon _{\rm b}'(k) = g \rho _{0} + \epsilon _{\rm b}(k)$. Then, in the low-wave-number region, 
$\epsilon _{\rm b}'(k) \simeq g \rho _{0} $ is approximately satisfied, so that they derived the $-3/2$ power law in the spectrum 
 $C_{\rm w}$  for the macroscopic wave function. 

However, owing to the weak fluctuation, the macroscopic wave function $\psi$ can be expanded as 
\begin{eqnarray}
\psi \simeq \varphi_{0} + \varphi_{1} + i\frac{\Pi_{1}}{2\hbar \varphi _0}.  \label{comp1}
\end{eqnarray}
By applying the Fourier transformation to Eq. (\ref{comp1}), we obtain
\begin{eqnarray}
\bar{\psi}(\bm{k}) \simeq \varphi_{0} \delta(\bm{k}) + \bar{\varphi}_{1}(\bm{k}) + i\frac{\bar{\Pi}_{1}(\bm{k})}{2\hbar\varphi _0}.  
\end{eqnarray}
As a result, in the low-wave-number region, the approximation used in  weak WT theory gives
\begin{eqnarray}
\langle |\bar{\psi}(\bm{k})|^2 \rangle \simeq \frac{mc_{\rm s}}{\hbar^2 \varphi_0^2 k} \langle |B(\bm{k})|^2 \rangle,   
\end{eqnarray}
which obviously leads to the $-7/2$ power law in $C_{\rm w}$ because $\mathcal{E}_{\rm b}(k) \propto k\langle |B(\bm{k})|^2\rangle k^{2} \propto k^{-3/2}$.
Thus, the derivation of the $-3/2$ power law for $C_{\rm w}$ in the previous work \cite{Proment09} seems not to be correct.

\subsection{Experimental possibility of Kolmogorov turbulence in atomic BECs}
In atomic BEC experiments, observations can be made of the density distribution, which accesses the spectrum $C_{\rm d}$. 
This can be the key to confirming the presence of Kolmogorov turbulence; if we can observe the $-3/2$ power law 
in this spectrum, then Kolmogorov turbulence is present. 

We discuss  three points  related to observing the $-3/2$ power law in $C_{\rm d}$. 
The first point is that one must  take a wide scaling region; that is, one must prepare an atomic BEC that is much larger than the coherence length. 
The $-3/2$ power law appears in the low-wave-number region $k < k_{\rm b}$, but, in the trapped system, there is a limit on the system
size of the order of the Thomas-Fermi radius $R_{\rm TF}$. This scale can be the lower limit of the scaling region for the $-3/2$ 
power law. Thus, the radius $R_{\rm TF}$ should be at least a factor of 10  larger than $2\pi /k_{\rm b}$.

The second point concerns how to excite the system weakly in the low-wave-number region. 
The $-3/2$ power law is valid under the weak nonlinearity condition $|\phi| \ll 1$, so that  fluctuations of both density and phase 
should be very weak. In our calculation, the density fluctuation is about $5 \%,$ as shown in Fig. 1. 
However, this may be too weak to observe. 
We must investigate the spectrum dependence on the strength of the fluctuation, for instance, 
the condition at which the prediction of  weak WT theory is broken; this, however,  is a subject of future work. 
Also, it is necessary to excite a large-scale modulation because the $-3/2$ power law results from the direct energy cascade. 
Thus,  candidates for the excitation are considered to be  repulsive or attractive Gaussian potentials moving slower than 
the sound velocity or a shaking trapping potential with fixed frequency. However, the excitation mode induced by these methods 
should depend on size, strength, and the velocity of the Gaussian potential or frequency and displacement of the shaking, etc. 
Understanding the excitation methods is essential for  observing Kolmogorov turbulence in atomic BECs and  should be numerically investigated hereafter. 

The third point is that one must take surface effects into consideration. In a three-dimensional system, we can obtain the density distribution integrated 
along the incident direction of the probe light, so that this distribution contains  information of the surface profile as well as the bulk. Since our result should be applicable to 
the bulk of the trapped BEC, the surface effect may alter the power law of Eq. (\ref{w_d_p_rel4}). 
If the system size is much larger than the damping length of the surface wave into the bulk \cite{PS}, this power law is expected to be relevant. 

Finally, we comment on  turbulence theory beyond  weak WT theory. 
Recently, Yoshida and Arimitsu have studied  strong turbulence with the spectral closure approximation, deriving 
the some power exponents \cite{Yoshida}. These power exponents have not been confirmed numerically. 
These power laws may be another key to confirmation of  Kolomogorov turbulence in  atomic BECs because, in this theory, 
the fluctuation is large and it is easy to observe.

\subsection{An inconsistency of our analytical calculation}
In Sec. III and Appendix A, we apply weak WT theory to Eqs. (\ref{density_equ}) and (\ref{Bogo_equ}), obtaining Eq. (\ref{kin_equ1}). 
However, this derivation has an inconsistency for the perturbative expansion in Eqs. (\ref{GP_condensate}) and (\ref{GP_fluctuation2}). 
In this expansion, we keep the second order of the fluctuation $\bar{\phi}$ in both equations, which is not consistent because the right-hand sides of Eqs. (\ref{GP_condensate}) and (\ref{GP_fluctuation2}) are the first and second order of smallness, respectively. 
To do a reliable expansion, we must take the third order of $\bar{\phi}$ in Eq. (\ref{GP_fluctuation2}), but this calculation is very difficult. 

Our numerical calculation exhibits that the condensate dynamics is different from $\sqrt{\bar{\rho} _{0} } {\rm exp}(-i\mu t/ \hbar)$, 
so that we must take the nonlinear dynamics for $\psi_0$. In this paper, although our calculation contains the inconsistency for the perturbative expansion, we treat the coupled nonlinear dynamics for the condensate and the fluctuation. Thus, our theory for this coupled dynamics is insufficient. The construction of the improved theory is the future work. In spite of this inconsistency, the analytical results exhibit good agreement with the numerical results. 

As another possibility, our numerical calculation seems not to be substantially satisfied with the weak nonlinear condition. However, if we use the weaker fluctuation, the computational time for the formation of the power law is much longer, which cannot be confirmed at present. 

\section{Conclusion}
We have theoretically and numerically studied  Bogoliubov-wave turbulence in a uniform BEC at zero temperature by using the GP equation. Our application of weak WT theory to the GP equation leads to the three power exponents $-5/2$, $-7/2$, and $-3/2,$ 
corresponding to the spectra for the Bogoliubov-wave distribution $C_{\rm b}$, the macroscopic wave function $C_{\rm w}$, and 
the density distribution $C_{\rm d}$. By numerically calculating the GP equation, we have confirmed the good agreement 
with these power exponents. 
In our analytical and numerical calculation, we found that taking account of the condensate dynamics coupled with 
the fluctuation, an aspect not treated in previous works \cite{Dyachenko92,Zakharov05,Proment09}, was important. 

\section*{Acknowledgments}
The authors thank S. Nazarenko for pointing out the importance of the dimension in the acoustic and Bogoliubov-wave turbulence. 
K.F. was supported by a Grant-in-Aid for JSPS Fellows
(Grant No. 262524). MT was supported by JSPS KAKENHI
Grant No. 26400366 and MEXT KAKENHI “Fluctuation $\&$ Structure” Grant No. 26103526.

\renewcommand{\theequation}{A\arabic{equation}}
\setcounter{equation}{0}
\section*{Appendix A: Application of weak WT theory}
We apply weak WT theory in \cite{wt2} to Eqs. (\ref{density_equ}) and (\ref{Bogo_equ}), deriving Eq. (\ref{kin_equ1}). 

In this theory, the separation of time scale is important. There are two time scales, which are a linear time scale $T_{\rm L} \sim 2 \pi \hbar/\epsilon_{\rm b}$ 
and a nonlinear one $T_{\rm NL}$. In weak WT, the nonlinear term is very weak, which leads to the long nonlinear time scale $T_{\rm NL} \gg T_{\rm L}$.
In weak WT theory, we make use of this time-scale separation, filtering out the fast linear oscillation.  
For this purpose, let us introduce auxiliary intermediate time $T$ satisfying $T_{\rm NL} \gg T \gg T_{\rm L}$. 
In the following, we solve Eqs. (\ref{density_equ}) and (\ref{Bogo_equ}) with a formal perturbative expansion, finding the canonical variables $b(\bm{k})$ at $t=T$. 

Before the perturbative expansion, we rewrite Eqs. (\ref{density_equ}) and (\ref{Bogo_equ}) with interaction variables $c(\bm{k})$ defined by 
\begin{eqnarray}
b(\bm{k}) = \epsilon c(\bm{k}) {\rm e}^{-i\alpha(k)}, 
\end{eqnarray}
\begin{eqnarray}
\alpha(k) = \frac{1}{\hbar} \int _{0}^{t} \epsilon _{\rm b}(k,t_1) dt_1,  
\end{eqnarray}
where we introduce the small parameter $\epsilon$ in order to elucidate the power of smallness of each term in the perturbative expansion. 
Then, Eqs. (\ref{density_equ}) and (\ref{Bogo_equ}) become
\begin{eqnarray}
i \hbar \frac{\partial}{\partial t} c(\bm{k}) = \epsilon \sum _{\bm{k}_1,\bm{k}_2} \Big[ \delta(\bm{k}-\bm{k}_1-\bm{k}_2) V(\bm{k},\bm{k}_1,\bm{k}_2) \nonumber \\
\times \Delta_1(\bm{k},\bm{k}_1,\bm{k}_2) c(\bm{k}_1) c(\bm{k}_2)  \nonumber \\ 
+ 2\delta(\bm{k}_1-\bm{k}_2-\bm{k}) V(\bm{k}_1,\bm{k}_2,\bm{k}) \nonumber \\
\times \Delta_2(\bm{k}_1,\bm{k}_2,\bm{k}) c(\bm{k}_1) c^* (\bm{k}_2)  \Big], \label{ap1}
\end{eqnarray}
\begin{eqnarray}
\frac{\partial}{\partial t} \rho_0 = \epsilon^2\frac{g\rho_0^2}{i\hbar} A  \label{ap2} 
\end{eqnarray}
with $\Delta_1(\bm{k},\bm{k}_1,\bm{k}_2) = {\rm exp}[i( \alpha(k)-\alpha(k_1)-\alpha(k_2) )]$ and $\Delta_2(\bm{k}_1,\bm{k}_2,\bm{k}) = {\rm exp}[-i( \alpha(k_1)-\alpha(k_2)-\alpha(k) )]$. We do not show the expression of $A$ because it is not important. In this calculation, as noted in Sec. III C, the terms with $W$ are neglected. 

We calculate the canonical variable $c(\bm{k})$ and the condensate density $\rho_0$ at $t=T$ up to the second order of $\epsilon$ by using the perturbative expansion:
\begin{eqnarray}
c(\bm{k}) = c^{(0)}(\bm{k}) + \epsilon c^{(1)}(\bm{k}) + \epsilon^2 c^{(2)}(\bm{k}) + \cdot\cdot\cdot ,   \label{ap3} 
\end{eqnarray}
\begin{eqnarray}
\rho_0 = \rho^{(0)}_0 + \epsilon \rho^{(1)}_0 + \epsilon^2 \rho^{(2)}_0 + \cdot\cdot\cdot  .  \label{ap4} 
\end{eqnarray}
From Eqs. (\ref{ap2}) and (\ref{ap4}), the expansion of $\rho_0$ is given by 
\begin{eqnarray}
\rho_0(T) = \rho_0(0)  + \epsilon^2 \rho^{(2)}_0(T) + \cdot\cdot\cdot  .  \label{ap5} 
\end{eqnarray}
The function $F(\rho_0)$ such as $V$, $\Delta_1$, and $\Delta_2$ has $\rho_0$ as a variable, being expanded in the following:
\begin{eqnarray}
F(\rho_0( T)) \simeq F(\rho_0(0)) + \frac{\partial F(\rho_0(0))}{\partial \rho_0(0)} \epsilon^2 \rho^{(2)}_0(T).  \label{ap6} 
\end{eqnarray}
Thus, the second order of the condensate density $\rho_0^{(2)}$ can be neglected in Eq. (\ref{ap1}) because such terms lead to the third order of $\epsilon$ in Eq. (\ref{ap1}). Therefore, the time dependence of $\rho_0$ vanishes in the application of weak WT theory, so that we can use the conventional derivation of the kinetic equation in \cite{wt2}, 
obtaining Eq. (\ref{kin_equ1}).
 
\renewcommand{\theequation}{B\arabic{equation}}
\setcounter{equation}{0}
\section*{Appendix B: Derivation of Eq. (\ref{Interaction4})}
The derivation of Eq. (\ref{Interaction4}) is described. 
In the low-wave-number region, the momentum and energy resonant conditions of the Bogoliubov waves give the particular class of the wave number, 
where the wave vectors are nearly parallel to each other because of the linear dispersion \cite{wt1, Fal, Lvov97}.
Then, the resonant condition $\epsilon_{\rm b}(k_1) = \epsilon_{\rm b}(k_2) +\epsilon_{\rm b}(|\bm{k}_1-\bm{k}_2|)$  becomes
\begin{eqnarray}
k_1-k_2 + \frac{1}{2k_{\rm b}^2} (k^3_1 - k_2^3)  \simeq |\bm{k}_1-\bm{k}_2| +  \frac{1}{2k_{\rm b}^2}|\bm{k}_1-\bm{k}_2|^3 , 
\end{eqnarray}
\begin{eqnarray}
 |\bm{k}_1-\bm{k}_2| \simeq \sqrt{(k_1-k_2)^2 + k_1k_2 \theta^2} , 
\end{eqnarray}
where the angle $\theta$ formed by the two vectors $\bm{k}_1$ and $\bm{k}_2$ is small.
Taking the second order of the small quantities, we obtain the angle $\theta \simeq \sqrt{3}(k_1-k_2)/k_{\rm b}$ \cite{wt1}, 
which gives 
\begin{eqnarray}
 |\bm{k}_1-\bm{k}_2| &\simeq& (k_1-k_2) \sqrt{1 + \frac{3k_1k_2}{k^{2}_{\rm b}}} \nonumber \\
 &\simeq& (k_1-k_2) \big( 1 + \frac{3k_1k_2}{2k_{\rm b}^2} \big).   
\end{eqnarray}
Here we use the condition $k_1>k_2$ because of the resonant condition.
Thus, the effective interaction between the Bogoliubov waves can be calculated by $V(\bm{k}_1,\bm{k}_2, \bm{k}_1-\bm{k}_2)$ of Eq. (\ref{Interaction3}).  
As a result, we obtain
\begin{eqnarray}
V_{\rm  b}(\bm{k}_1, \bm{k}_2, \bm{k}_1-\bm{k}_2) = \frac{3g\rho_0}{2k_{\rm b}^{3/2}} \sqrt{k_1k_2(k_1-k_2)}.  
\end{eqnarray}

\renewcommand{\theequation}{C\arabic{equation}}
\setcounter{equation}{0}
\section*{Appendix C: Derivation of Eq. (\ref{bogkin1})}
We show the derivation of Eq. (\ref{bogkin1}) from Eq. (\ref{kin_equ1}). 
We assume the isotropy of $n(\bm{k})$ in the weak WT, integrating the $\bm{k}_2$ part and taking the leading terms, which leads to
\begin{eqnarray}
\frac{\partial}{\partial t} n(k) = I(k) , 
\end{eqnarray}
\begin{widetext}
\begin{eqnarray}
I(k) = \frac{4\pi^2}{\hbar c_{\rm s}} \int_0 ^{\infty}  d k_1 \int _0^{\pi} {\rm sin } \theta _1 d\theta_1 \Biggl[   \frac{2(k-k_1)k_1}{k} |V_{\rm b}(k, k_1,|\bm{k}-\bm{k}_1|)|^2 \theta_{\rm H}(k-k_1) \delta _{\rm d} ( \theta_1^2 -  \Theta_1^2) f(k,k_1,k-k_1) \nonumber \\
 - \frac{2(k_1-k)k_1}{k} |V_{\rm b}(k_1, k,|\bm{k}_1-\bm{k}|)|^2  \theta_{\rm H}(k_1-k) \delta _{\rm d} ( \theta_1^2 -  \Theta_1^2) f(k_1,k,k_1-k) \nonumber \\ 
 - \frac{2(k_1+k)k_1}{k} |V_{\rm b}(|\bm{k}_1+\bm{k}|,k_1, k,)|^2 \delta _{\rm d} ( \theta_1^2 -  \Theta_2^2) f(k+k_1,k_1,k)   \Biggl],
\end{eqnarray} 
\end{widetext}
with $f(k,k_1,k_2) =  n(k_1)n(k_2) - n(k_2)n(k)- n(k)n(k_1)$, $\Theta _1 = \sqrt{3}|k-k_1|/k_{\rm b}$, and $\Theta _2 = \sqrt{3}(k+k_1)/k_{\rm b}$.
Finally, we can perform the integration by substitution with $y=\theta_1^{2}$ and ${\rm sin} (\Theta_j) \simeq  \Theta _j$ ($j=1,2$), obtaining Eq. (\ref{bogkin1}).  

\renewcommand{\theequation}{D\arabic{equation}}
\setcounter{equation}{0}
\section*{Appendix D: Locality of the power law}
The locality condition of the power law is derived. 
In this derivation, we consider the convergence condition of collisional integral in  Eq. (\ref{bogkin1}). 

The power law is assumed to be $n(k) \propto k^{-m}$ with the positive power exponent $m$. 
Then, from Eq. (\ref{bogkin1}), we find that the collisional integral may diverge at three points $k_1=0,k,\infty$. 
The point $k_1=k$ means the $k_2=0$, so that the convergence condition at $k_1=k$ is the same as that at $k_1=0$ because of the exchange symmetry 
($\bm{k_1}  \leftrightarrow \bm{k}_2$) of Eq. (\ref{kin_equ1}).  
Thus, in the following, the convergence conditions at $k_1=0,\infty$ are described. 

In the large $k_1$ region, the dangerous terms in the collisional integral are the second and third ones of Eq. (\ref{bogkin1}), which is approximately expressed by 
\begin{eqnarray}
&2&\int ^{\infty} dk_1 \Bigl[   k_1^4 n(k) \Big( n(k_1+k)-n(k_1-k) \Big) \Bigl] \nonumber \\
&\simeq& 4\int ^{\infty} dk_1 kk_1^{4}n(k) \frac{\partial n(k_1)}{\partial k_1} \nonumber \\
&\propto& \int ^{\infty} dk_1 k_1^{3-m}.
\end{eqnarray}
Thus, the convergence condition at $k_1=\infty$ is $4<m$. 

On the other hand, in the small $k_1$ region, the dangerous terms appear in the first and third ones of Eq. (\ref{bogkin1}). 
These terms become
\begin{eqnarray}
& &\int _0 dk_1 \Bigl[   k_1^2 n(k_1) \Big( n(k+k_1)+n(k-k_1)-2n(k) \Big) \Bigl] \nonumber \\
&\simeq& \int _0 dk_1 k_1^{4}n(k_1) \frac{\partial^2 n(k)}{\partial k^2} \nonumber \\
&\propto& \int_0 dk_1 k_1^{4-m}, 
\end{eqnarray}
which shows that the convergence condition at $k_1=0$ is $m<5$.

In summary, the convergence condition of the collisional integral is $4<m<5$. 
In the Bogoliubov-wave turbulence, the power exponent is $9/2$, so that this power law is local.

\end{document}